\documentstyle{article}

\newcommand{\mathbf}{\bf}

\begin{document}

\begin{center}
{\huge\bf  On The Quantum Theory of Hall Effect }
\end{center}

\vspace{1cm}
\begin{center}
{\large\bf
F.GHABOUSSI}\\
\end{center}

\begin{center}
\begin{minipage}{8cm}
Department of Physics, University of Konstanz\\
P.O. Box 5560, D 78434 Konstanz, Germany\\
E-mail: ghabousi@kaluza.physik.uni-konstanz.de
\end{minipage}
\end{center}

\vspace{1cm}

\begin{center}
{\large{\bf Abstract}}
\end{center}

\begin{center}
\begin{minipage}{12cm}
We discuss a model of both classical and integer quantum  
Hall-effect which is based on a semi-classical  
Schroedinger-Chern-Simons-action, where the Ohm-equations result as  
equations of motion. The quantization of the classical  
Chern-Simons-part of action under typical quantum Hall conditions  
results in the quantized Hall conductivity. We show further that the  
classical Hall-effect is described by a theory which arises as the  
classical limit of a theory of quantum Hall-effect. The model  
explains also the preference and the domain of the edge currents on  
the boundary of samples.

\end{minipage}
\end{center}

\newpage

{\large{\bf Introduction and summary}}

Recently, we discussed a model of the integer quantum Hall-effect  
(IQHE) \cite{allgemein} according to which the quantization of  
Hall-conductivity should result from the quantum electrodynamics in  
2+1-dimensions \cite{mein one}.
In this semi classical Schroedinger-Chern-Simons-model the  
Hall-conductivity $\sigma_H$ appears as the normalization parameter  
of the Chern-Simons-action \cite{N}. Furthermore, we assumed there  
according to the experimental results of QHE a vanishing  
longitudinal conductivity $\sigma_L$\cite{allgemein}. Thereafter,  
the Ohm-equations of IQHE with quantized $\sigma_H$ are obtained as  
the equations of motion from the Schroedinger-Chern-Simons-action  
with quantized electromagnetic potentials \cite{mein one}.

Here we discuss a more general model for both classical Hall-effect  
(CHE) and IQHE, where the related Ohm-equations result as equations  
of motion also from a Schroedinger-Chern-Simons-action functional.  
Thereafter, the quantum Hall conditions \cite{kk} cause the  
transition of the Hall-system into the quantum regime, where the  
necessary quantization of electromagnetic potentials results in the  
quantized $\sigma_H$ in the absence of $\sigma_L$. It is a model of  
non-intercating charge carriers for IQHE with a semi-classical  
Schroedinger-Chern-Simons-action functional, hence not the  
Schroedinger-term which represents the charge carriers system but  
only the Chern-Simons-term which represents the dynamics of the  
almost pure gauge potentials is quantized \cite{mein one}. Thus, a  
second quantization of the Schroedinger-term in our model which  
corresponds to the interacting particle system should result after  
solution of question of the ground state, in a FQHE model similar to  
the known models \cite{N}.

\medskip
Our model is based on the following stand point on the theory of  
Hall-effects that because there are both CHE and QHE (IQHE and  
FQHE), thus the theory of the QHE must be the quantization of the  
"classical" theory of the CHE \cite{N}. Furthermore a rigorous  
quantization of a system requires the knowledge of its action  
functional. Accordingly, we have to construct first a "classical"  
action for the CHE, wherefrom the resulting equations of motion must  
explain the CHE behaviour. On the other hand the "classical"  
Ohm-equations \cite{allgemein} are the only equations which describe  
the CHE. Thus, the action which should describe the CHE has to  
result in the Ohm-equations as its equations of motion. This  
interpretation of the Ohm-equations as the equations of motion which  
must result directly from an action functional is a new element of  
our stand point. In all other models the Ohm-equations are  
considered as a given relation in the sense of "material" or  
"phenomenological" equations \cite{F}.

On the other hand, in view of the well known fact that these  
Ohm-equations are semi-classical relations with Schroedinger-typ  
current densities for electrons, the desired action for CHE should  
be also of the semi-classical typ as it is performed in our model.  
Then, the canonical quatization of the classical part of this action  
for the case of non-interacting electrons must result in the  
quantum theory of the IQHE and also in the quantized  
Hall-conductivity according to the IQHE.

\bigskip

To investigate the relation between QHE and CHE, let us analyse  
first the Ohm-equations for QHE and CHE \cite{allgemein}.
These are given by:

\begin{equation}
j_m = \sigma_H \epsilon_{nm} E_n \;\;\; ,\;\;\; \epsilon_{mn} =  
-\,\epsilon_{nm} = 1\;\;\;; m,n = 1,2\;\;\;,
\end{equation}
\label{ohmq}

for QHE, where $\sigma_H = {\displaystyle\frac{en}{B}}$ becomes  
quantized in the units of ${\displaystyle\frac{e^2}{h}}$. Here $n$  
is the global surface density of the charge carriers ("electrons")  
which we call electrons and $B:=B_3$ is the applied magnetic field  
\cite{erklar}.

On the other hand, the Ohm-equations for CHE are given by:

\begin{equation}
j_m = \sigma_H \epsilon_{nm} E_n + \sigma_L E_m
\end{equation}
\label{cohm}

with $\sigma_L = {\displaystyle\frac{\sigma_0}{1 +  
(\omega_c\tau)^2}}$ and $\sigma_H =  
{\displaystyle\frac{\sigma_0(\omega_c\tau)}{1 + (\omega_c\tau)^2}}$,  
where $\sigma_0 = {\displaystyle\frac{e^2 n \tau}{\mu}}$, $\omega_c  
:= {\displaystyle\frac{eB}{\mu}}$, $\tau$ and $\mu$ are the mean  
free time and the mass of electrons [1] \cite{lasst}.

\bigskip
The key observation is that according to quantum mechanics  
\cite{landau} the current density of electrons in a {\it magnetic  
field} without spin term and with $C = 1$ is given by (a): $j_m:=  
{\displaystyle\frac{ie\hbar}{2\mu}} [ (\partial_m \psi^*)\psi -  
\psi^* (\partial_m\psi)] - {\displaystyle\frac{e^2}{\mu}  
A_m\psi^*\psi}$, whereas the current density of electrons in the  
limit $B\rightarrow 0$, i. e. for $\omega_c\tau \ll 1$ should be  
given by (b): $j_m:= {\displaystyle\frac{ie\hbar}{2\mu}} [  
(\partial_m \psi^*)\psi - \psi^* (\partial_m\psi)]$. Both obeying  
the continuity equation \cite{landau}.

We deduce that the relation (a) is valid in the integer quantum  
Hall-regime $(\omega_c\tau\gg 1)$ where the external magnetic field  
is large, whereas the relation (b) is valid in the classical  
Hall-regime $(\omega_c\tau\ll 1)$ where the same external field is  
small or absent.

\medskip
The semi-classical Schroedinger-Chern-simons-action functional in  
$2+1$-dimensions is the only action from which we can obtain the  
mentioned Ohm-equations (1) and (2) as the equations of motion (see  
below) \cite{erkme}, where the $\sigma_H$ plays the role of  
normalization parameter of the classical Chern-Simons-action.

\medskip
To see the relation of the quantization of Hall-system with the  
empirical quantum behavior under the typical Quantum Hall-conditions  
\cite{kk} let us recall that in a strong magnetic field the  
Hall-conductivity $\sigma_H$ becomes {\it small} according to its  
definition which is given above \cite{allgemein}. Precizely, in the  
quantum Hall-limit, i. e. $\omega_c\tau\gg 1$ the $\sigma_H$ and  
$\sigma_L$ should be considered according to their definitions which  
is given above of the orders $(\omega_c\tau)^{-1}$ and  
$(\omega_c\tau)^{-2}$ respectively, i. e. $\sigma_H \ll 1$ and  
$\sigma_L\ll\sigma_H$ or $\sigma_L\rightarrow 0$. Moreover, in this  
limit the Hall-conductivity is given by $\sigma_H =  
{\displaystyle{\frac{ne}{B}}}$ so that for small n and large  
$B_{external}$ the $\sigma_H$ becomes considerablly small. Thus, if  
we consider in our model, $\sigma_H$ as the normalization parameter  
of the Chern-Simons-action $S_{C-S}$ \cite{N} and quantize this  
action according to the Schroedinger representation \cite{mein one}:

\begin{equation}
\Psi_{(C-S)}(A) \propto e^{\displaystyle{i\frac{\sigma_H  
S_{C-S}}{\hbar}}} \;\;\;,
\end{equation}
\label{altact}

the $\sigma_H S_{C-S}$ becomes also small for relevant $S_{C-S}$  
actions in view of the above mentioned smallness of $\sigma_H$.  
Therefore, for small $\sigma_H S_{C-S}$, i. e. precisely for those  
$\sigma_H S_{C-S}$, which are comparable with $\hbar$, the quantum  
behaviour of action becomes dominant \cite{feyn} and we meet the  
integer quantum Hall-regime manifested by IQHE. Moreover, in this  
quantum limit the $\sigma_L$ becomes, as mentioned above, {\it very  
small} tending to zero as it is expected in the QHE.

Conversely, if the magnetic field is not strong, i. e. for  
$\omega_c\tau \ll 1$, $\sigma_H$ and $\sigma_H S_{C-S}$ become large  
or $\sigma_H S_{C-S}\gg\hbar$ and we meet the classical regime,  
where the quantum fluctuations of the action are compensated  
\cite{feyn} and the original quantum theory reduces to its classical  
limit which is the theory of CHE. In this classical limit $\sigma_L  
\approx \sigma_0$, thus both typ of conductivities are no more  
small but of considerable magnitudes, since they are also present in  
the Ohm-equations of the CHE (2). We avoid to discuss here the  
typical FQH-conditions including the high mobility of particles in  
view of the fact that we consider only the IQHE.

On the other hand, it is known that if one considers currents  
involved in the IQHE only as the boundary currents, then most of  
experimental data can be understood in a satisfactory manner  
\cite{kk}. It is a favour of the Chern-Simons-ansatz in a manifold  
with a spatial boundary that the boundary currents are the only  
allowed ones according to the constraints of the theory under the  
typical quantum Hall-conditions \cite{kk}.
Therefore, for construction of a theoretical model for both CHE and  
IQHE one is left with the Schroedinger-Chern-Simons-action, from  
which we know already that it results, at least, in the  
Ohm-equations for CHE and IQHE as the equations of motion \cite{mein  
one}.

\bigskip
{\large{\bf The Chern-Simons-Action for Classical and Quantum  
Hall-Effect}}

The general action from which the Ohm-equations of CHE and IQHE  
((2) and (1)) can be obtained as the equations of motion is the  
following Schroedinger-Chern-Simons-action defined on the  
2+1-dimensional manifold $M = \Sigma\times\mathbf R$.

\begin{equation}
S = \frac{1}{8\pi}\int dt \int_{\Sigma} \psi^* [ i\hbar\partial_t -  
\frac{1}{2\mu} (-i\hbar\partial_m - e A_m )^2 - eA_0 ]\psi + h.c. -  
\frac{\sigma_H}{8\pi}
\int_M\epsilon^{\alpha\beta\gamma}A_{\alpha}\partial_{\beta}A_{\gamma}  
\;,
\end{equation}
\label{action}

where $A_{\alpha} (x_m,t)$ is still the classical electromagnetic  
potential which remains classical in the classical Hall-regime but  
must be quantized in the quantum Hall-regime. Furthermore,  
${\{\alpha,\beta,\gamma}\} = {\{0,1,2}\}$ and everywhere is $C = 1$,  
$\partial_m = {\displaystyle\frac{\partial}{\partial x_m}}$ and we  
consider (in accordance with the experimental arrangements of the  
QHE) that the $\Sigma$ has a boundary. Furthermore, as already  
mentioned the Schroedinger-term represents the mechanics of the  
non-interacting particle system, whereas the Chern-Simons-term  
represents the dynamics of the coupled electromagnetic potentials.

Obviously, we use the $\sigma_H$ as the locally constant  
normalization parameter of the Chern-Simons-action. It is justified  
to do so, because $\sigma_H$ can be considered as a dimensionless  
and locally constant quantity in $2+1$-dimensios also in view of its  
well known topological or global character \cite{allgemein}  
\cite{N} \cite{diff}. Moreover, we suppressed the spin term within  
the usual Schroedinger-action for "electron" in a magnetic field in  
view of the well known fact about QHE that the spin degenerecy is  
not essential for the IQHE \cite{allgemein} .

In view of the gauge freedom of $A_m$ we choose the gauge fixing  
condition $A_0 = 0$ to retain the true degrees of freedom of the  
electromagnetic fields in the action (4).
Thereafter, the action reduces to the following one:

\begin{equation}
\frac{1}{8\pi}\int dt \int_{\Sigma} \psi^* [ i\hbar\partial_t -  
\frac{1}{2\mu} (-i\hbar\partial_m - e A_m )^2 ]\psi + h.c.-  
\frac{\sigma_H}{8\pi}\int dt \int_{\Sigma} \epsilon^{mn} \dot{A}_m  
A_n \;\;,
\end{equation}
\label{act2}

The equations of motion for classical $A_m$ potentials which result  
from this action are

\begin{equation}
j_m - {\displaystyle\frac{e^2 n}{\mu}} A_m =  \sigma_H  
\epsilon^{nm} \dot{A}_n \;\;\;,
\end{equation}
\label{eqmot}

where we used according to $\omega_c\tau \ll 1$ in the classical  
regime the corressponding definition (b) $j_m:=  
{\displaystyle\frac{ie\hbar}{2\mu}} [ (\partial_m \psi^*)\psi -  
\psi^* (\partial_m\psi)]$.

We introduce the gauge $A_m = E_m \tau$ in (6) which is more  
appropriate for the case of low magnetic fields, i. e. precisely it  
is appropriate for the classical Hall-regime with $\omega_c\tau\ll  
1$ \cite{spaet}.
It is equivalent to the relaxation time approximation which is the  
usual approach in this case \cite{relax}. Substituting $ A_m = E_m  
\tau$ in (6) we obtain the desired Ohm-equations for CHE

\begin{equation}
j_m = \sigma_L E_m+ \sigma_H \epsilon_{nm} E_n\;\;\;,
\end{equation}
\label{ohmc}

where we used $\sigma_L \approx\sigma_0$ according to  
$\omega_v\tau\ll 1$.

Thus, we obtained the Ohm-equations of the CHE as the equations of  
motion from the action (4) in the classical Hall-regime,  
consistently, according to $\omega_c\tau \ll 1$.

\bigskip
The quantization of the action (5) under the typical IQH-conditions  
\cite{kk}, i. e. in the limit $\omega_c\tau\gg 1$ results then in  
the action which is responcible for the Ohm-equations of IQHE, where  
one must use obviously the defition (a) for the current density in  
the quantum Hall-regime according to $\omega_c\tau\gg 1$.

Recalling our previous analysis we like to mention that the quantum  
regime of
Hall-effect is related in double sense to the strong exterior  
magnetic field which is applied on the two dimensional electronic  
systems: In the limit $\omega_c\tau\gg 1$ the $\sigma_H$ and  
$\sigma_L$ should be considered theoretically of the orther  
$(\omega_c\tau)^{-1}$ and $(\omega_c\tau)^{-2}$ respectively, i. e.  
$\sigma_H$ becomes small and $\sigma_L$ tends to zero, as it is  
confirmed by experiments \cite{kk}. On the other hand, under typical  
quantum Hall conditions where the number or the density of  
electrons is small \cite{kk} the $\sigma_H$ and $\sigma_H S_{C-S}$  
become more smaller and so the last one becomes comparable with  
$\hbar$ which results in the integer quantization of $\sigma_H$ as  
it is also confirmed by experiments \cite{kk}.

In other words, the $\omega_c\tau \gg 1$ limit together with small  
$n$ corresponds with the quantum regime \cite{kk} where $\sigma_H  
S_{C-S}$ becomes comparable with $\hbar$, whereas the $\omega_c\tau  
\ll 1$ limit together with $n$ around the usual electronic density  
in metals corresponds with the classical limit where the action  
$\sigma_H S_{C-S}\gg \hbar$.
Therefore, for large magnetic fields and small density of electrons  
which are the typical quantum Hall conditions the two dimensinal  
Hall-system is in the IQHE-regime \cite{kk} which is described by  
the same action (4) or (5) after gauge fixing:

\begin{equation}
\frac{1}{8\pi}\int dt \int_{\Sigma} \psi^* [ i\hbar\partial_t -  
\frac{1}{2\mu} (-i\hbar\partial_m - e A_m )^2 ]\psi + h.c.-  
\frac{\sigma_H}{8\pi}\int dt \int_{\Sigma} \epsilon^{mn} \dot{A}_m  
A_n \;\;\;,
\end{equation}
\label{act3}

but in view of $\sigma_H S_{C-S}\approx \hbar$ with $A_m$  
potentials now obeying the usual quantization algebra \cite{witten  
and}
\begin{equation}
\left[\hat{A}_m(x_l,t) \,, \, \hat{A}_n(y_l,t) \right] =  
\frac{4{\pi}i\hbar}{\sigma_H}\,\,\epsilon_{mn}\delta^2( X - Y )    
\;\; ; \: X,Y \in \Sigma \;\;,
\end{equation}
\label{comutat}

which can be red off directly from the Chern-Simons-action in (8).
It means that $\hat{A}_m := {\displaystyle\frac{i\partial}{\partial  
A_n}}$ which is the usual polarization of the ${\{A_m}\}$ phase  
space.

\bigskip
However, for practical use it is convinient to introduce the  
Schroedinger representation $\Psi(A) \propto  
e^{\displaystyle{\frac{i}{\hbar}\sigma_HS_{C-S}}}$ of the   
Chern-Simons-action

\begin{equation}
- \frac{\sigma_H}{8\pi}\int dt \int_{\Sigma} \epsilon^{mn}  
\dot{A}_m A_n \;\;,
\end{equation}
\label{act4}

after its quantization according to (9), hence $\Psi(A)$ must  
fullfil the relation (9) in the sense of its eigen functions.

To obtain $\Psi(A)$ we use the method introduced in a previous work  
on IQHE \cite{mein one}. It is based on the representation of the  
state functions $\Psi(A)$ in terms of the eigen states of the  
quantum orbital angular momentum.
For equivalent quantization of $S_{C-S}$ and its Schroedinger  
representations see \cite{witten and}.

\medskip
Introducing polar coordinates in the phase space described by the  
action (10), the quantum orbital angular momentum becomes $\hat{L} =  
-i\hbar\partial_{\phi}$ \cite{landau}. Thereafter, $\Psi(A)$ is  
given as the eigen states of the operator $\hat{L}$ by:

\begin{equation}
\Psi (A) = F(R)\, e^{{\;\displaystyle\frac{i}{\hbar} \sigma_H  
l\phi}} \;\;,
\end{equation}
\label{psi}

Here F(R) is an arbitrary function of R and $l = R^2$ is the value  
of angular momentum of the system which is a constant of motion  
according to the $SO(2)$ symmetry of the system. We normalize the  
constant $l = 1$.

Thus, the necessary single-valuedness of $\Psi(A)$ forces the $\sigma_H$
to be

\begin{equation}
 \sigma_H = 0, 1, 2, ..., N, ... \;\;; N\in\mathbf Z_+\;\;\;,
\end{equation}
\label{sigma}

where we restricted us to the positive values \cite{jak}.

\medskip
Recall that the normalization parameter of the $\Psi_{C-S}$ becomes  
allways quantized as integers in view of the single valuednes of  
$\Psi_{C-S}$ in its first quantization no matter what kind of  
quantization is performed \cite{witten and}.

Empirically it is the mentioned typical IQH-conditions \cite{kk}
which prepares the electrons, according to their density and  
mobility and the strength of the exterior magnetic fields, to be in  
IQHE situation (see also the conclusion).

\bigskip
The equations of motion for $A_m$ potentials which result from the  
quantized action (8) for the non-interacting system of charge  
carriers, according to (11)-(12) and using the corresponding  
definition (a) for the current density in magnetic fields, are:

\begin{equation}
j_m = \sigma_H\epsilon_{nm}E_n \;\;\;\;,
\end{equation}
\label{ohmnew}

which are the desired Ohm-equations with quantized $\sigma_H$.

\bigskip
It is obvious from the comparison between the quantized  
Chern-Simons-action in units of $\hbar$, i. e.  
${\displaystyle\frac{\sigma_H S}{\hbar}}$ and the  
Schroedinger-action in (8) that in the atomic units the $\sigma_H$  
should be considered in units of ${\displaystyle\frac{e^2}{h}}$,  
which is equivalent to a redefinition of the quantized  
$A_m$-potentials absorbing the coupling constant $e$.

Thus, we obtained the quantized Ohm-equations of IQHE as the  
equations of motion from the quantized  
Schroedinger-Chern-Simons-action.

\bigskip
To summerize the quantum and classical behavior in this model let  
us recapitulate the analysis of the integer quantum and classical  
Hall conditions:

If the Hall-system is prepared with $\omega_c\tau\gg 1$ and with  
small $n$, then the quantum modes of its action become dominant, but  
if it is prepared with $\omega_c\tau\ll 1$ and with $n$ around the  
density of CHE-samples then its classical modes become dominant.

The theoretical description of this situation is according to our  
model so that the general semi-classical action functional for both  
cases should be given by (4) where the Schroedinger-term remains the  
same in both cases in view of the non-interacting particles in  
IQHE. Then, the action (4) with quantized Chern-Simons-term  
describes the integer quantum Hall-regime, whereas the action (4)  
with classical Chern-Simons-term describes the classical  
Hall-regime.

In the first case the typical quantum Hall conditions, i. e.  
$\omega_c\tau\gg 1$, and small $n$ cause the smallness of $\sigma_H$  
so that $\sigma_H S_{C-S}$ becomes comparable with $\hbar$. Thus  
the quantum modes of the action
$\sigma_H S_{C-S}$ which are represented by $\Psi(A)$ become  
dominant requiring the quantization of $\sigma_H$. Since the total  
quantum action results in the "quantum" Ohm-equations with  
integrally quantized $\sigma_H$ and vanishing $\sigma_L$ as it is  
shown above.

In the second case the action is of the order $\sigma_H  
S_{C-S}\gg\hbar$, therefore the classical limit of  
Chern-Simons-action, i. e. the classical Chern-Simons-action becomes  
dominant. Then the total action reduces to the semi-classical  
Chern-Simons-Schroedinger-action which describe the semi-classical  
theory of the CHE. Since it results in the "classical" Ohm-equations  
as it is shown above.

\medskip
Thus, the theory of CHE, i. e. its action arises as the classical  
limit from the quantum action of IQHE.

\newpage

{\large{\bf The Edge Currents in QHE}}

Obviously, the motion of system which is described by the action  
(8) together with the quantization relations (9(-(12) is constrained  
by
the constraint:

\begin{equation}
     -\sigma_H \epsilon^{mn}\partial_m A_n = e\psi^*\psi \,\,,
\end{equation}
\label{contraint}

with $ e\psi^*\psi:= j_0$.

If we integrate the relation (14) over the sample surface and  
consider $B :=\epsilon_{nm} \partial_m A_n$ as a constant field  
strength, then we obtain the well known relation between the  
Hall-conductivity and the magnetic field, namely

\begin{equation}
{\displaystyle\sigma_H = \frac{ne}{B}} \;\;\;,
\end{equation}
\label{contraint2}

where $n = (a)^{-1}\int da (\psi^*\psi)$ is the global density of  
charge carriers and $a$ is the area of sample.

Recall, that the relation (15) is conforme with the general  
definition of
$\sigma_H$ in the limit $\omega_c\tau\gg 1$ \cite{xx}.

\bigskip
However, the constraint (14) influences the motion of the  
IQHE-system in a way which is known from the experimental results of  
IQHE.

To see this let us note first some of main experimental features of  
IQHE rewieved from \cite{kk}:

1. Most of IQHE-data can be understood in a satisfactory manner if  
one reduces the involved currents to the edge currents.

2. The typical IQHE-regime is related to $B\gg$ and small $n$.

3. Under integer quantum Hall conditions the edge of Hall-systems  
are chractrized by
the $n\rightarrow 0$.

4. For the large current densities the IQHE can not be simply  
described by the edge currents located on the boundary, whereas the  
low currents are transported by the edge channels.

All these features of IQHE can be understood if we take into  
account the constraint (14).

Recall that, in view of the Ohm-equations the currents are  
restricted to those regions where the $A_m$-potentials are allowed  
to exist. Thus, the question of the edge currents is related with  
the questions of the regions where the $A_m$-potentials are defined.  
Moreover, according to the constraint (14) the potential $A_m$  
becomes pure gauge potential with vanishing field strength if  
$n\rightarrow 0$.

This is the case if one has to do with samples with small $n$ under  
the large $B$ for example on the edges of quantum Hall-system. Thus  
under these circumstances we should replace the costraint (14) by  
the following one

\begin{equation}
           \epsilon^{mn}\partial_m A_n \approx 0 \;\;\;,
\end{equation}
\label{cosnt3}

for systems under quantum Hall conditions \cite{kk}.
Thereafter, the $A_m$-potentials become pure gauge potentials, i.e.  
$A_m \approx ig^{-1} \partial_m g$, where g is an element of the  
U(1)-gauge group. Recall however that this is a local relation in  
quantum mechanics, therefore 1.) it should be valid only within the  
limit of uncertainty relations and 2.) a locally pure gauge  
potential has the well known geometric, i. e. globally well defined  
and observable effects in quantum mechanics \cite{nnn}.

On the other hand, the constraint tensor $\epsilon_{mn}\partial_m  
A_n$ generates a gauge transformation $A_m^\prime = A_m +\partial_m  
\lambda$ in the phase space of the $A_m$-potentials \cite{witten  
and}.
Therefore, according to the constraint (16) one must identify  
$A_m^\prime = A_m$ everywhere in the phase space.
Furthermore, if as in our case the $\Sigma$ possess a boundary we  
must  choose boundary conditions for $A_m$ and $\lambda$ on the  
boundary. We choose free boundary conditions for $A_m$ but $\lambda  
= 0$ on the boundary. A reason for this choise is that the  
Chern-Simons-action is not invariant under gauge transformations  
that do not vanish on the boundary \cite{witten and}.

Accordingly, it must be required that $A_m^\prime = A_m$ for any  
$\lambda$ which vanishes on the boundary $\partial\Sigma$. The only  
pure $A_m$ gauge potentials which obey this additional condition are  
those restricted to be defined only on the boundary \cite{witten  
and}. In other words, the only $A_m$-potentials obeying both  
restrictions caused by the constraint (16) are those restricted to  
exists on the boundary region of $\Sigma$. Thereafter, the currents  
$j_m$ should be considered also to be restricted to the boundary  
region of $\Sigma$, i. e. to the so called edge currents.  
Accordingly, under quantum Hall conditions \cite{kk} the edge  
currents are the prefered ones.

\medskip
It is importent to mention that if we consider this restrictions of  
the potentials and currents to the boundary or to the edge of  
Hall-system "quantum mechanically", then there is an uncertainty of  
the position of currents, or so to say there is an uncertainty of  
the "quantum mechanical" edge $\Delta({\partial\Sigma})$ in view of  
the Heisenberg's uncertainty relations. Thus, if we consider the  
uncertainty of momentum equal to $(2m \Delta E)^{\frac{1}{2}}$ with  
$\Delta E = E_{n+1} - E_{n} = {\displaystyle{\frac{\hbar  
\omega_c}{2}}}$ the uncertainty of the mentioned edge or the width  
of the current's orbit is given by $\Delta X =  
({\displaystyle{\frac{\hbar}{eB}}})^{\frac{1}{2}}$ which is the  
magnetic length $l_B$. Since, the edge current is according to its  
empirical definition the current which flows, in the ideal case,  
close to the edge within the length scale of the magnetic length  
\cite{kk}. Moreover, this circumstance shows also that the  
constraint (16) should be fullfield within the uncertainty dictated  
by the energy-time uncertainty relation. Since the $\Delta E \propto  
\Delta B$ in the Landau-levels \cite{landau}.

\medskip
On the other hand, if $n >$ for large transport currents the right  
hand side of the constraint (14) and thereby also the field strength  
in (14) is obviously non-vanishing and the IQHE breaks down as  
manifested by early experiments \cite{kk}.

\bigskip
{\large{\bf Conclusion}}
:This was a model of IQHE based on the non-interacting system of  
charge carriers coupeled on an electromagnetic potential in  
$2+1$-dimensions. There are strong hints that the FQHE which is  
belived to be a many particle effect, i. e. of interacting  
particles, should results from the second quantization of the  
Schroedinger-field of charge carriers involved in an action similar  
to one which is used in this model \cite{N}. Hence, the conformity  
of our model for IQHE with an erlear model of FQHE \cite{N} is a  
hint about the possibility that, if one consider a proper  
modification of our model for the case of interacting charge  
carriers, then after the second quantization of the  
Schroedinger-term in our action for the interacting ("many  
particle") system one should arrive in a theory of FQHE.
However, this is possible if one can solve the problem of ground  
state of interacting particles in such models \cite{N}.  We discuss  
the second quantization of our model and the resulting fractionality  
elsewhere \cite{under}.

\bigskip

\bigskip

\bigskip
Footnotes and references


\begin{thebibliography}{100}

\bibitem{allgemein}
For a general review on QHE and its experimental setting see:

[1a] R.E. Prange and S.M. Girvin, ed., The quantum Hall effect,  
Graduate Texts in Contemporary Physics (Springer, New York, 1987);

[1b] A.H. Macdonald, ed., Quantum Hall effect: A Perspective,  
Perspectives in Condensed Matter Physics (Kluver Academic  
Publishers, 1989)

[1c] G. Morandi, The role of Topology in Classical and Quantum  
Physics, Lecture Notes in Physics m7 (Springer, New York 1992)

[1d] M. Janssen, et al, ed., J. Hajdu, Introduction to the Theory  
of the Integer Quantum Hall effect (VCH-verlag, Weinheim, New York,  
1994)

\bibitem{mein one}
F. Ghaboussi, On the Integer Quantum Hall Effect, KN-UNI-preprint-95-1;
A Model of the Integer Quantum Hall Effect, KN-UNI-preprint-95-2,  
submited for publication.

\bibitem{N}
The use of $\sigma_H$ as normalization parameter of  
Chern-Simons-action is in accordance with the use of similar  
parameters in interacting system of particles which become  
afterwards proportional to $\sigma_H$ in FQHE models: G. W. Semenof,  
Phs. Rev. Lett. 61, 517, (1988); S. C. Zhang, T. H. Hanssson and S.  
Kivelson, Phys. Rev. Lett. 62, 82 (1989). Recall that it is  
expected that non-interacting particles in quantum Hall-samples  
result under proper conditions in the IQHE, whereas the interacting  
particle systems should be responcible for the fractional QHE  
(FQHE). In the last case it seems that depending on the theoretical  
treatment of the question of the ground state one is lead to one of  
the above mentioned models.

\bibitem{kk}
K. von Klitzing, Physica B 204 (1995) 111-116;
R. Knott, W. Dietsche, K. von Klitzing, K. Eberl and K. Ploog,  
Semicond. Sci. Technol. 10 (1995) 117-126

\bibitem{F}
For a different model, where only the Ohm-equations of the IQHE but  
not that of CHE are drived as equations of motion see: J.  
Fr{\"o}hlich, T. Kerler, Nuc. Phys. B354 (1991) 369-417.


\bibitem{erklar}
Precisely, the total magnetic field acting on the Hall system  
described by the
Schroedinger-Chern-Simons-action (9) is given by $B_{total} :=  
B_{external} +
B(A_m)$ with $B_{external}\gg B(A_m)$, where $B_{external}$ is the  
external homogenous strong magnetic field applied on the system. The  
$B(A_m)$ is the magnetic field arised from the dynamics of  
$A_m$-potentials which is also responcible for the electric fields  
$E_m$. The $B(A_m)$ is usually so small that $\omega_c\tau\ll 1$ and  
so its influence on the conductivity is contained already in what  
is known under the classical Hall-effect. Since, to achive magnetic  
influence of the quantum Hall-type one needs strong magnetic fields  
as those used in QHE-experiments (see refs. [1] and \cite{kk}).

\bibitem{lasst}
Recall also that relation (2) can be obtained from relation (1) by  
an infinitesimal SO(2)-transformation in the ${E_m}$- or in the  
${A_m}$-space. The infinitesimal angle $\delta\chi =  
{\displaystyle{\frac{\sigma_L}{\sigma_H}}}$
becomes almost zero in the quantum Hall-regime.

\bibitem{landau}
See L.D.Landau, E.M.Lifschitz, III Vol.

\bibitem{erkme}
See also the Ref. \cite{mein one}.

\bibitem{feyn}
R. P. Feynman and A. R. Hibbs, Quantum mechanics and Path Integrals  
(McGraw-Hill 1965)


\bibitem{diff}
This means that $d\sigma_H = 0$. Furthermore, recall also that both  
charge carier dencity n in two dimensions and the B-field are of  
dimension $L^{-2}$. Thus, the $\sigma_H =  
{\displaystyle\frac{en}{B}}$ becomes dimensionless. For further  
arguments in favour of the local constancy of $\sigma_H$ see the  
paper quoted in Ref. \cite{F}.

\bibitem{spaet}
Recall that in presence of magnetic fields the well known  
Landau-gauge is given by $A_m = B x_n\epsilon_{mn}$ (see ref.  
\cite{landau}).

\bibitem{relax}
J. Callaway, Energy band Theory (Academic Press 1964). Recall also  
that the relaxation time $\tau$ is indeed introduced in this  
approximation to calculate the electric conductivity from the  
Ohm-equations. One can show that the usual relaxation time  
approximation results in the approximation $\Delta A_m = E_m\Delta t  
\approx E_m \tau$, where the defining vanishing average velocity   
$\bar{V} = 0$ is given according to the operator $\hat{V}_m =  
\hat{P}_m - e \hat{A}_m$.

\bibitem{witten and} E. Witten, Cumm. Math. Phys. 121 (1989) 351-399;
R. Jakiw, in: Physics, Geometry, and Topology, Nato ASI Series,  
ed., H.C. Lee (Plenum Press, New York, 1990);
G.V. Dunne, R. Jackiw, and C Treugenberger, Ann. Phys. 194 (1989)  
197-223;
G.V.Dunne and C. Treugenberger, Mod. Phys. Lett. A Vol4 (1989) 1635-1644

\bibitem{jak}
See for other methodes of quantizations Ref. \cite {witten and}.  
The fractional quantization of the normalization parameter should be  
a result of multivaluedness of the wave function of the electrons  
in the Schroedinger-term in its second quantization  which is  
related with the interacting electrons (see the models quoted in  
Ref. \cite{N}). It is well known that the mentioned properties of  
electrons like the mobility and also the strength of exterior  
magnetic field has different values for the FQHE-samples  
\cite{allgemein}.

\bibitem{xx}
Recall also that $\sigma_H$ becomes $\sigma_H = {\displaystyle  
\frac{ne}{B}}$ only in the quantum Hall-limit, whereas in the  
classical Hall-limit it is given by
$\sigma_H = \sigma_0 \omega_c\tau$.

\bibitem{nnn}
It is this discrepancy between the local and global properties of  
pure gauge potential which makes a classical or a either  
semi-classical understanding of QHE difficult. The quantum  
mechanical, i. e. the global or invariant character of pure gauge  
potential is given by its line integral which is the phase of wave  
function and results in the flux quantization. Furthermore we mean  
here always a pure gauge potential of the electromagnetic or  
U(1)-type in a multiply connected region.

\bibitem{under}
Under preparation.
\end{thebibliography}
\end{document}